\documentstyle[aps,preprint,tighten,epsfig]{revtex}
\def\beq{\begin{equation}}
\def\eeq{\end{equation}}
\def\beqa{\begin{eqnarray}}
\def\eeqa{\end{eqnarray}}
\def\MeV{\nobreak\,\mbox{MeV}}
\def\GeV{\nobreak\,\mbox{GeV}}
\def\keV{\nobreak\,\mbox{keV}}

\def\pli{p^\prime}
\def\mli{{M^\prime}^2}
\begin{document}
\title{\sc  $D^*D\pi$ and $B^*B\pi$ form factors from QCD Sum Rules}
\author {F.S. Navarra, M. Nielsen \\
\vspace{0.3cm}
{\it Instituto de F\'{\i}sica, Universidade de S\~{a}o Paulo, } \\
{\it C.P. 66318,  05389-970 S\~{a}o Paulo, SP, Brazil}\\
\vspace{0.3cm}
M.E. Bracco, M. Chiapparini and C. L. Schat\\
\vspace{0.3cm}
{\it Instituto de F\'{\i}sica,  Universidade do Estado do Rio de Janeiro, } \\ 
{\it Rua S\~ao Francisco Xavier 524, Maracan\~a,  20559-900, Rio de Janeiro, 
RJ, Brazil}}
\maketitle
\vspace{1cm}

\begin{abstract}
The $H^*H\pi$ form factor for $ H = B $ and $D $ mesons 
 is evaluated in a QCD sum rule calculation. 
We study the Borel sum rule for the three
point function of two pseudoscalar and one vector meson currents
up to order four in the operator product expansion. The double Borel transform
is performed with respect to the heavy meson momenta. 
We discuss the momentum dependence of the form factors and two different
approaches to extract the $H^*H\pi$ coupling constant.
\\
PACS numbers 14.40.Lb,~~14.40.Nd,~~12.38.Lg,~~11.55.Hx
\\

\end{abstract}

\vspace{1cm}
The coupling of the pion to the  heavy  mesons ($g_{B^*B\pi}$ and 
$g_{D^*D\pi}$)
is related to the form factor at zero pionic momentum and its precise
value has been often needed in phenomenology. In particular, the
$g_{D^*D\pi}$ coupling is needed in the context of quark gluon plasma (QGP) 
physics.
Suppression of charmonium production in heavy ion  collisions is one of 
the signatures of QGP formation \cite{ms}. Therefore  a precise 
evaluation of the background, i.e.,  conventional 
$J/\psi$ absorption by co-moving  pions and $\rho$ mesons 
\cite{ko}, is of fundamental 
importance. Since pions are so abundant in a dense nuclear environment, 
the reactions $\pi \, + \, J/\psi \, \rightarrow \, D \, + \, \overline{D^*}$ 
(and consequently the coupling $g_{D^*D\pi}$)
are of special relevance \cite{mamu}.

In the case of $g_{D^*D\pi}$, the ${D^*}^+\rightarrow D^0\pi^+$ decay is
observed experimentally. However, present data provide only an upper bound:
$g_{D^*D\pi}\leq 21 $ \cite{pdg}. For $g_{B^*B\pi}$, there cannot be a direct
experimental indication because there is no phase space for the
$B^{*}\rightarrow B\pi$ decay. Recently, a direct preliminary determination of 
$g_{B^*B\pi}$ on the lattice has been attempted \cite{ukqcd}.

The $D^*D\pi$ and $B^*B\pi$ couplings have been studied by several authors 
using different approaches of the QCD sum rules (QCDSR): two point function
combined with soft pion techniques \cite{col,ele}, light cone sum rules
\cite{bel,col2}, light cone sum rules including perturbative corrections 
\cite{kho}, sum rules in a external field \cite{gro}, double
momentum sum rules \cite{dn}. Unfortunately, the numerical results from these 
calculations may differ by almost a factor two.

In this work we use the three-point function approach to evaluate the 
$D^*D\pi$ and $B^*B\pi$ form factors and coupling constants. The advantage
of using the three-point function approach with a double Borel transformation
compared with the two-point function with a single Borel transformation
is the elimination of the  terms associated with the pole-continuum 
transitions \cite{bel,io1}.

The three-point function associated with a $H^*H\pi$ vertex, where $H$ and
$H^*$ are respectively the lowest pseudoscalar and vector heavy
mesons, is given by
\begin{equation}
\Gamma_\mu(p,\pli)=\int d^4x \, d^4y \, \langle 0|T\{j(x)
j_5(y)j^\dagger_\mu(0)\}|0\rangle  
\, e^{ip^\prime.x} \, e^{-i(\pli-p).y}\; , 
\label{cor}
\end{equation}
where $j=i\bar{Q}\gamma_5 u$, $j_5=i\bar{u} \gamma_5d $ and
$ j^\dagger_\mu=\bar{d}\gamma_\mu Q$ are the interpolating fields for $H$, 
$\pi^-$ and $H^*$ respectively with $u$, $d$ and $Q$ being the up, down, and 
heavy quark fields.

The phenomenological side of the vertex function, $\Gamma_\mu(p,p^\prime)$,
is obtained by the consideration of $H$ and $H^*$ state contribution to
the matrix element in Eq.~(\ref{cor}):
\beqa
\Gamma_\mu^{(phen)}(p,\pli)&=&{1\over p^2-m_{H^*}^2}{1\over{\pli}^2-m_{H}^2}
\langle 0|j|H(\pli)\rangle\times 
\nonumber \\*[7.2pt]
&&\langle H(\pli)|j_5|H^*(p,\epsilon)\rangle\langle H^*(p,\epsilon)|
j^\dagger_\mu|0\rangle + \mbox{higher resonances}\; .
\label{phe1}
\eeqa

The matrix element of the pseudoscalar element, $j_5$, defines the
vertex form factor $g_{H^*H\pi}(q^2)$:
\beq
\langle H(\pli)|j_5|H^*(p,\epsilon)\rangle={f_\pi m_\pi^2\over m_u+m_d}{
g_{H^*H\pi}(q^2)\over q^2-m_\pi^2}q_\nu\epsilon^\nu\; ,
\label{for}
\eeq
where $q=\pli-p$, $f_\pi$ is the pion decay constant and
$\epsilon^\nu$ is the polarization of the vector meson. The vacuum to meson 
transition amplitudes appearing in Eq.~(\ref{phe1}) are given in terms of
the corresponding meson decay constants $f_H$ and $f_{H^*}$ by
\beq
\langle 0|j|H(\pli)\rangle={m_H^2f_H\over m_Q}\;,
\label{fh}
\eeq
and
\beq
\langle H^*(p,\epsilon)|j^\dagger_\mu|0\rangle=m_{H^*}f_{H^*}\epsilon^*_\mu
\; .
\label{fh*}
\eeq
Therefore, using Eqs.~(\ref{for}), (\ref{fh}) and (\ref{fh*}) in
Eq.~(\ref{phe1}) we get
\beqa
\Gamma_\mu^{(phen)}(p,\pli)&=&C_{HH^*}{g_{H^*H\pi}(q^2)\over q^2-m_\pi^2}
{1\over p^2-m_{H^*}^2}{1\over{\pli}^2-m_{H}^2}\times
\nonumber \\*[7.2pt]
&&\left(-\pli_\mu+{m_{H^*}^2+m_H^2-q^2\over2m_{H^*}^2}p_\mu\right)
+ \mbox{higher resonances}\; ,
\label{phen}
\eeqa
where
\beq
C_{HH^*}={m_H^2m_{H^*}m_\pi^2f_Hf_{H^*}f_\pi\over (m_u+m_d) m_Q}\; .
\label{chh}
\eeq
The contribution of higher resonances and continuum in Eq.~(\ref{phen})
will be taken into account as usual in the standard form of 
ref.~\cite{io2}.

The QCD side, or theoretical side, of the vertex function is evaluated by
performing Wilson's operator product expansion (OPE) of the operator
in Eq.~(\ref{cor}). Writing $\Gamma_\mu$ in terms of the invariant
amplitudes:
\beq
\Gamma_\mu(p,\pli)=\Gamma_1(p^2,{\pli}^2,q^2)p_\mu + \Gamma_2(p^2,{\pli}^2,q^2)
\pli_\mu\;,
\eeq
we can write a double dispersion relation for each one of the invariant
amplitudes $\Gamma_i\,(i=1,2)$, over the virtualities $p^2$ and ${\pli}^2$
holding $Q^2=-q^2$ fixed:
\beq
\Gamma_i(p^2,{\pli}^2,Q^2)=-{1\over4\pi^2}\int_{m_Q^2}^\infty ds
\int_{m_Q^2}^\infty du {\rho_i(s,u,Q^2)\over(s-p^2)(u-{\pli}^2)}\;,
\label{dis}
\eeq
where $\rho_i(s,u,Q^2)$ equals the double discontinuity of the amplitude
$\Gamma_i(p^2,{\pli}^2,Q^2)$ on the cuts $m_Q^2\leq s\leq\infty$,
$m_Q^2\leq u\leq\infty$, which can be evaluated using Cutkosky's rules
\cite{io2,cut}.

Finally we perform a double Borel transformation \cite{io2} in both variables
$P^2=-p^2$ and ${P^\prime}^2=-{\pli}^2$ and equate the two representations
described above. We get one sum rule for each invariant function. In the
$p_\mu$ structure:
\beqa
-C_{HH^*}{m_{H^*}^2+m_H^2+Q^2\over2m_{H^*}^2}
{g_{H^*H\pi}(q^2)\over Q^2+m_\pi^2}e^{-m_{H^*}^2/M^2}
e^{-m_{H}^2/{\mli}}&=&
-{1\over4\pi^2}\int_{m_Q^2}^{s_0} ds
\int_{m_Q^2}^{u_0} du \left[\right.
\nonumber \\*[7.2pt]
&&\left.\rho_1(s,u,Q^2)e^{-s/M^2} e^{-u/\mli}\right]\;,
\label{srp}
\eeqa
and in the $\pli_\mu$ structure:
\beq
C_{HH^*}{g_{H^*H\pi}(q^2)\over Q^2+m_\pi^2}e^{-m_{H^*}^2/M^2}
e^{-m_{H}^2/{\mli}}=
-{1\over4\pi^2}\int_{m_Q^2}^{s_0} ds
\int_{m_Q^2}^{u_0} du \left[
\rho_2(s,u,Q^2)e^{-s/M^2} e^{-u/\mli}\right]\;,
\label{srpl}
\eeq
where $s_0$ and $u_0$ are the continuum thresholds for the $H^*$ and $H$ 
mesons respectively, which are, in general, taken from the mass sum rules.
The two Borel masses $M^2$ and $\mli$ are, in principle, independent
and they should vary in the vicinity of the corresponding meson masses:
$m_{H^*}^2$ and $m_H^2$ respectively. Since for heavy mesons $m_H$ and
$m_{H^*}$ are very close, many authors use $ M^2=\mli$ \cite{bel,kho,gro}.
To allow for different values of $ M^2$ and $\mli$ we take them proportional
to  the respective meson masses, which leads us 
to study the sum rule as a function of $M^2$ at a fixed ratio 
\beq
{M^2\over\mli}={m_{H^*}^2\over m_H^2}\;.
\label{bo}
\eeq

We will consider diagrams up to dimension four which include the perturbative 
diagram and the gluon condensate. The quark condensate term does not contribute
since it depends only on one external momentum and, therefore, it is
eliminated by the double Borel transformation. Higher dimension condensates
are strongly suppressed in the case of heavy quarks 
\cite{col,ele,bel,col2,gro,dn}.
The double discontinuity of the perturbative contribution reads:
\beq
\rho_1^{(pert)}(s,u,Q^2)=-{3Q^2u(2m_Q^2-s-u-Q^2)\over[(s+u+Q^2)^2-4su]^{3/2}}
\; ,
\eeq
\beq
\rho_2^{(pert)}(s,u,Q^2)={3Q^2[m_Q^2(s+u+Q^2)-2su]\over[(s+u+Q^2)^2-4su]^{3/2}}
\; ,
\eeq
and the integration limit condition is
\beq
(s-m_Q^2)(u-m_Q^2)\geq Q^2m_Q^2\; .
\eeq

In this paper we focus on the structure $p_{\mu}$ which we found to be the more
stable one. For consistency we use in our analysis the QCDSR expressions for 
the decay constants up to dimension four in lowest order 
of $\alpha_s$:
\beq
f_H^2={3m_Q^2\over 8\pi^2m_H^4}\int_{m_Q^2}^{u_0}du 
{(u-m_Q^2)^2\over u}e^{(m_H^2-u)/\mli}\,-\, {m_Q^3\over m_H^4}
\langle\bar{q}q\rangle e^{(m_H^2-m_Q^2)/\mli}\; ,\label{fhr}
\eeq
\beq
f_{H^*}^2={1\over 8\pi^2m_{H^*}^2}\int_{m_Q^2}^{s_0}ds {(s-m_Q^2)^2\over s}
\left(2+{m_Q^2\over s}\right)e^{(m_{H^*}^2-s)/M^2}
\,-\,{ m_Q\over m_{H^*}^2}\langle\bar{q}q\rangle e^{(m_{H^*}^2-m_Q^2)/M^2}
\; ,\label{fhs}
\eeq
where we have omitted the  numerically insignificant contribution of the
gluon condensate.

The parameter values used in all calculations are $m_u+m_d=14\,\MeV$, 
$m_c=1.5\,
\GeV$, $m_b=4.7\,\GeV$, $m_\pi=140\,\MeV$, $m_D=1.87\,\GeV$, $m_{D^*}=2.01\,
\GeV$, $m_B=5.28\,\GeV$, $m_{B^*}=5.33\,\GeV$, $f_\pi=131.5\,\MeV$,
$\langle\overline{q}q\rangle\,=\,-(0.23)^3\,\GeV^3$, $\langle g^2G^2\rangle 
=0.5\,\GeV^4$. We parametrize the continuum thresholds as
\beq
s_0=(m_{H^*}+\Delta_s)^2\;,\label{s0}
\eeq
and 
\beq
u_0=(m_{H}+\Delta_u)^2\;.\label{u0}
\eeq
 The values of $u_0$ and $s_0$ are, in general, extracted from the 
two-point
function sum rules for $f_H$ and $f_{H*}$ in Eqs.~(\ref{fhr}) and (\ref{fhs}).
Using the Borel region $2 \leq M^2\leq 5 \GeV^2$ (for the $D^*$ and  $D$ 
mesons) and $10 \leq M^2 \leq 25 \GeV^2$ (for the $B^*$, and $B$ mesons) 
we found a good stability for $f_H$ and $f_{H*}$ with 
 $\Delta_s=\Delta_u\sim0.5\GeV$, in agreement with the results in 
ref.~\cite{bel}. We have checked that bigger values for $\Delta_{s(u)}$, of 
order of 1 GeV, lead to unstable results for $f_H$ and $f_{H*}$, in the 
case of the sum rules Eqs.~(\ref{fhr}) and (\ref{fhs}). 
In our study we will allow for a small 
variation in  $\Delta_s$ and $\Delta_u$ to test the sensitivity of our
results to the continuum contribution.

We first discuss the $D^*D\pi$ form factor. In Fig.~1 we
show the behavior of the perturbative and gluon condensate contributions
to the form factor $g_{D^*D\pi}(Q^2)$ at $Q^2=1\,\GeV$ as a function
of the Borel mass $M^2$ using $\Delta_s$ and $\Delta_u$ given in Eqs.
(\ref{s0}) and (\ref{u0}) equal to $0.5\,\GeV$. We can see that, in the case 
of the form factor, the gluon condensate is not negligible and it helps the 
stability of the curve, as a function of $M^2$, providing a rather stable
plateau for $M^2\geq3\,\GeV^2$. The behavior of the curve for other $Q^2$
and continuum treshold values is similar. Fixing $M^2=3.5\,\GeV^2$ we show, 
in Fig.~2,
the momentum dependence of the form factor (dots). Since the present approach 
cannot be used at $Q^2=0$, to extract the $g_{D^*D\pi}$
coupling from the form factor we need to extrapolate the curve to $Q^2=0$
(in the approximation $m_\pi^2=0$). In order to do this extrapolation we fit 
the
QCD sum rule results (dots) with an analytical expression. We tried to fit
our results with a monopole form, since this is very often used 
for form factors, but the fit is very poor. We obtained  good fits using  
both  the gaussian form 
\beq
g_{H^*H\pi}(Q^2)= g_{H^*H\pi} \,  e^{-(Q^2+m_\pi^2)^2/\Gamma^4}
\label{ga}
\eeq
and a curve of the form
\beq
g_{H^*H\pi}(Q^2)=g_{H^*H\pi}{1+(a/\Lambda)^4\over1+(a/\Lambda)^4
e^{(Q^2+m_\pi^2)^2/\Lambda^4}}\;.
\label{ff}
\eeq
In Fig.~2 we show that the $Q^2$ dependence of the 
form factor, represented by the dots, can be
well reproduced by the parametrization in Eqs.~(\ref{ga}) (dashed line) and
(\ref{ff}) (solid line). The value of the  parameters in Eqs.~(\ref{ga}) 
and (\ref{ff}) are given 
in Table I for two different values of the continuum threshold.
\vskip 5mm
\begin{center}
\begin{tabular}{|c|c|c|c|c|}
\hline
$\Delta_s=\Delta_u\,(\GeV)$ & $g_{D^*D\pi}$ & $\Lambda\,(\GeV)$&$a\,(\GeV)$&
$\Gamma \, (\GeV)$\\
\hline\hline
0.5 & 5.3 & 1.66 & 1.90 & -\\
0.6 & 6.0 & 1.89 & 3.05 & -\\
0.5 & 5.7 & - & - & 1.74\\
0.6 & 6.1 & - & - & 1.92\\
\hline 
\end{tabular}
\end{center}
\begin{center}
\bf{TABLE I:} {\small Values of the parameters in Eqs.~(\protect\ref{ga}) and 
(\protect\ref{ff})
which reproduce the QCDSR results for $g_{D^*D\pi}(Q^2)$,
for two different values of
the continuum thresholds in Eqs.~(\protect\ref{s0}) and (\protect\ref{u0}).}
\end{center}
\vskip5mm
In view of the uncertainties involved, the  results obtained with the two 
parametrizations 
are consistent with each other, the systematic error being of the order of 
$10 \%$.

In refs.~\cite{bel,bbd} it was found that the form factor in the
semileptonic decay $H\rightarrow\pi l\bar{\nu}$, which is also normalized by 
the $H^*H\pi$ coupling constant, can be well approximated
by a monopole form factor. In the case of
the $H\rightarrow\pi l\bar{\nu}$ form factor, a vector dominance approximation
gives a phenomenological explanation for a pole fit at $q^2=m_{H^*}^2$,
which is not the case of the form factor studied here.
It is important to notice that here the dispersion relation is written in  
terms of the two heavy meson momenta, while in the case of 
semileptonic decay the dispersion relation is a function of the
$H$ and $\pi$ momenta. Therefore, our form factor is a function of the
pion momentum, exhibiting a peak at the pion pole $Q^2=0$.
 
To test if our fit gives a good extrapolation to $Q^2=0$ we can write a
sum rule, based on the three-point function Eq.~(\ref{cor}), but valid only
at $Q^2=0$, as suggested in \cite{rry} for the pion-nucleon coupling
constant. This method was also applied to the nucleon-hyperon-kaon
coupling constant \cite{ccl,bnn} and to the nucleon$-\Lambda_c-D$ coupling
constant \cite{nn}. It consists in neglecting the pion mass in the denominator
of Eq.~(\ref{phen}) and working at $Q^2=0$, making a single Borel 
transformation
to both $P^2={P^\prime}^2\rightarrow M^2$.

As discussed in the introduction, the problem of doing a single Borel 
transformation is the fact that the single pole contribution, associated
with the $N\rightarrow N^*$ transition, is not suppressed \cite{col,bel,io1}.
In ref.~\cite{io1} it was explicitly shown that the
pole-continuum transition has a different behavior as a function of the
Borel mass as compared with the double pole contribution and continuum
contribution: it grows with $M^2$ as compared with the double pole 
contribution. Therefore, the single pole contribution can be taken into 
account through the introduction of a parameter $A$, in the phenomenological 
side of the sum rule \cite{bel,io1,bnn}. Thus, neglecting $m_\pi^2$ in the
denominator of Eq.~(\ref{phen}) and doing a single Borel transform in
$P^2={P^\prime}^2$, we get for the structure $p_\mu$
\beq
\tilde{\Gamma}_1^{(phen)}(M^2,Q^2)=-{C_{H^*H}\over2m_H^2Q^2}{m_H^2 + 
m_{H^*}^2+Q^2
\over m_{H^*}^2-m_H^2}\left(e^{- m_H^2/M^2}-e^{- m_{H^*}^2/M^2}\right)
(g_{H^*H\pi}+AM^2)\; ,
\label{0phe}
\eeq
where $C_{H^*H}$ in given in Eq.~(\ref{chh}) with $f_H$ and $f_{H^*}$ given
by Eqs.~(\ref{fhr}) and (\ref{fhs}).

On the OPE side only terms proportional to $1/Q^2$ will contribute to the sum 
rule. Therefore, up to dimension four the only diagram that contributes is
the quark condensate given by
\beq
\tilde{\Gamma}_1^{<\bar{q}q>}(M^2,Q^2)={2m_Q\langle\overline{q}q\rangle\over
Q^2}\,e^{-m_Q^2/M^2}\; .
\label{qbq}
\eeq

Equating Eqs.~(\ref{0phe}) and (\ref{qbq}) and taking $Q^2=0$ we obtain the 
sum rule for $g_{H^*H\pi}+AM^2$, where $A$ denotes the contribution from the 
unknown single poles terms. It is interesting to point out that in the
limit $m_H^2+m_{H^*}^2=2m_{H^*}^2$, the sum rule obtained in the $\pli_\mu$
structure coincides with the sum rule in the $p_\mu$ structure. In Fig.~3
we show, for $\Delta_s=\Delta_u=0.5\,\GeV$, the QCDSR results for 
$g_{D^*D\pi}+AM^2$
as a function of $M^2$ (dots) from where we see that, in the Borel region
$2\leq M^2\leq5\,\GeV^2$, they follow a straight line. The value of the 
coupling constant is obtained by the extrapolation of the line to $M^2=0$
\cite{io1}. Fitting the QCDSR results to a straight line we get
\beq
g_{D^*D\pi}\simeq5.4\;,
\eeq
in excellent agreement with the values obtained with the extrapolation of the
form factor to $Q^2=0$, given in Table I.

It is reassuring that both methods, with completely different OPE sides
and Borel transformation approaches, give the same value for the coupling 
constant. 

In the case of $B^*B\pi$ vertex, we show in Fig.~4, for $\Delta_s=\Delta_u
=0.5\,\GeV$, the $Q^2=0$ sum rule results for $g_{B^*B\pi}+AM^2$ (dots)
as a function of $M^2$. It also follows a straight line in the Borel
region $10\leq M^2\leq25\,\GeV^2$, and the extrapolation to $M^2=0$ gives
\beq
g_{B^*B\pi}\simeq10.6\;.
\eeq

In Fig.~6 we show the QCDSR result for the perturbative and gluon condensate
contributions to the form factor $g_{B^*B\pi}(Q^2)$ at $Q^2=2\,\GeV^2$ as
a function of $M^2$ using $\Delta_s=\Delta_u=0.5\,\GeV$. In this case the
gluon condensate is very small but it still goes in the right direction of 
providing a stable plateau for $M^2\geq15\,\GeV^2$. Fixing $M^2=17\,\GeV^2$
we show, in Fig.~6, the $Q^2$ behavior of the form factor (dots). The dots
can still be well fitted   by Eq.~(\ref{ff}) (solid line). However, 
the fit with Eq.~(\ref{ga}) is not so good, as can be seen by the dashed 
line in Fig.~6. 
In Table II we give the
value of the parameters in Eqs.~(\ref{ga}) and (\ref{ff}) that reproduce our 
results for two  different choices of the continuum thresholds.

In this case the agreement of the two different approaches to extract
the coupling constant is not so good, but the numbers are still
compatible. One possible reason for that is the fact that for heavier quarks
the perturbative contribution (or hard physics) becomes more important,
as can be observed by the decrease of the importance of the gluon condensate 
in Fig.~5 as compared with  Fig.~1. Since in the sum rule given by Eqs.
(\ref{0phe}) and (\ref{qbq}) there is only soft physics information,
we expect  $\alpha_s$ corrections to the sum rule to be more
important in the case of $g_{B^*B\pi}(Q^2)$ than for $g_{D^*D\pi}(Q^2)$.
\vskip 5mm
\begin{center}
\begin{tabular}{|c|c|c|c|c|}
\hline
$\Delta_s=\Delta_u\,(\GeV)$ & $g_{B^*B\pi}$ & $\Lambda\,(\GeV)$&$a\,(\GeV)$ & 
$\Gamma \, (\GeV)$\\
\hline\hline
0.5 & 14.7 & 1.62 & 1.37 & -\\
0.6 & 16.3 & 1.81 & 1.67 & -\\
0.5 & 17.2 & - & - & 1.79\\
0.6 & 18.4 & - & - & 1.97\\
\hline 
\end{tabular}
\end{center}
\begin{center}
\bf{TABLE II:} {\small Values of the parameters in Eqs.~(\protect\ref{ga}) and 
(\protect\ref{ff})
which reproduce the QCDSR results for $g_{B^*B\pi}(Q^2)$,
for two different values of
the continuum thresholds in Eqs.~(\protect\ref{s0}) and (\protect\ref{u0}).}
\end{center}
\vskip5mm
Comparing Table I with Table II we see that the cut-offs are of 
the same order in the two vertices and are very hard. Concerning the 
parameter $a$,
it is smaller in the case of the $B^*B\pi$ vertex. This is because
of the fact that the form factor $g_{B^*B\pi}(Q^2)$ has a flatter peak around 
$Q^2=0$ than  $g_{D^*D\pi}(Q^2)$. This can be interpreted as an indication 
that the spatial extension of the vertex is smaller for $B^*B\pi$ than
for $D^*D\pi$. This is also the reason why the gaussian fit is not so good
in the case of the $B^*B\pi$ vertex, and leads to bigger values for the
coupling. It is interesting to notice that
our results for the coupling constants are completely consistent with the QCDSR
calculation of ref.~\cite{dn}.

As a final exercise,  we  use our result for $g_{B^*B\pi}$ to extract the
coupling constant $g$ which controls the interaction of the pion with
infinitely heavy fields in effective lagrangian approaches \cite{cas,sin}.
They are related by  \cite{col,ele,bel,col2,gro,dn,cas,sin}
\beq
g_{B^*B\pi}={2m_B\over f_\pi}g\;.
\eeq
The knowledge of $g$ is of great phenomenological value, since its strenght
is required in the analyzes of many electroweak processes \cite{cas}. 
Therefore, during the last years, a large number of theoretical papers has 
been devoted to the calculation of $g$. However, the variation of the
value obtained for $g$, even within a single class of models, turns out
to be quite large. For instance, using different quark models one obtains
$1/3\leq g\leq 1$ \cite{sin,dam} while  QCDSR calculations points in the
direction of small $g$, with a typical value in the range
$g\simeq0.13 - 0.35$ \cite{col,ele,bel,col2,gro,dn}.

Using the values for $g_{B^*B\pi}$ given in Table II we get, at order 
$\alpha_s=0$ 
\beq
g=0.13 - 0.23\;,
\eeq
therefore, we corroborate the overall conclusion drawn from different QCDSR 
calculations, that the coupling $g$ is small.

In conclusion, we extracted the $H^*H\pi$ coupling constant using two different
approaches of the QCDSR based on the three-point function. We have obtained 
for the coupling constants:
\beq
g_{D^*D\pi}=5.7\pm0.4\;,
\label{gd}
\eeq
\beq
g_{B^*B\pi}=14.5\pm3.9\;,
\eeq
where the errors reflect variations in the continuum thresholds,
different parametrizations of the form factors and the use of two different 
sum rules. There are still sources of errors in the values of the condensates
and in the choice of the Borel mass to extract the form factor, which were
not considered here. Therefore, the errors quoted are probably underestimated.

In Table III we present a compilation of the estimates of the coupling
constants $g_{D^*D\pi}$ and $g_{B^*B\pi}$  from distinct QCDSR 
calculations.
\vskip 5mm
\begin{center}
\begin{tabular}{|c|c|c|}
\hline
approach & $g_{D^*D\pi}$ &  $g_{B^*B\pi}$ \\
\hline\hline
this work & $5.7\pm 0.4$ & $14.5\pm3.9$\\
two-point function + soft pion techniques (2PFSP)\cite{col} & $9\pm2$ & 
$20\pm4$\\
2PFSP + perturbative corrections \cite{col} & $7\pm2$ & $15\pm4$\\ 
light cone sum rules (LCSR) \cite{bel} & $11\pm2$ & $28\pm6$ \\
LCSR + perturbative corrections \cite{kho} & $10.5\pm3$ & $22\pm9$\\
double momentum sum rule \cite{dn} & $6.3\pm1.9$ & $14\pm4$\\ 
\hline 
\end{tabular}
\end{center}
\begin{center}
\bf{TABLE III:} {\small Summary of QCDSR estimates for  $g_{D^*D\pi}$ and
 $g_{B^*B\pi}$.}
\end{center}
\vskip5mm

From this Table we see that our result is in a fair agreement with the
calculations in refs.~\cite{col,dn}, while LCSR calculations point to
bigger values for the coupling constants. This discrepancy has still to be 
solved.

The $D^*D\pi$ coupling is directly related with the $D^*\rightarrow D\pi$ 
decay width through
\beq
\Gamma({D^*}^-\rightarrow {\overline{D^0}} \pi^-)={g_{D^*D\pi}^2|
\vec{q}_\pi|^3\over24\pi m_{D^*}^2}\;.
\eeq
Using Eq.~(\ref{gd}) we get
\beq
\Gamma({D^*}^-\rightarrow {\overline{D^0}}\pi^-)=6.3\pm0.9\;\keV\;,
\eeq
which is much smaller then the current upper limit \cite{pdg}
$\Gamma({D^*}^-\rightarrow {\overline{D^0}}\pi^-)<89$ keV.

\vspace{1cm}
 
\underline{Acknowledgements}: 
This work has been supported by CNPq and FAPESP (under
project number 1998/2249-4). C.L.S. thanks FAPERJ for financial support.

\begin{figure} \label{fig1}
\begin{center}
\epsfysize=9.0cm
\epsffile{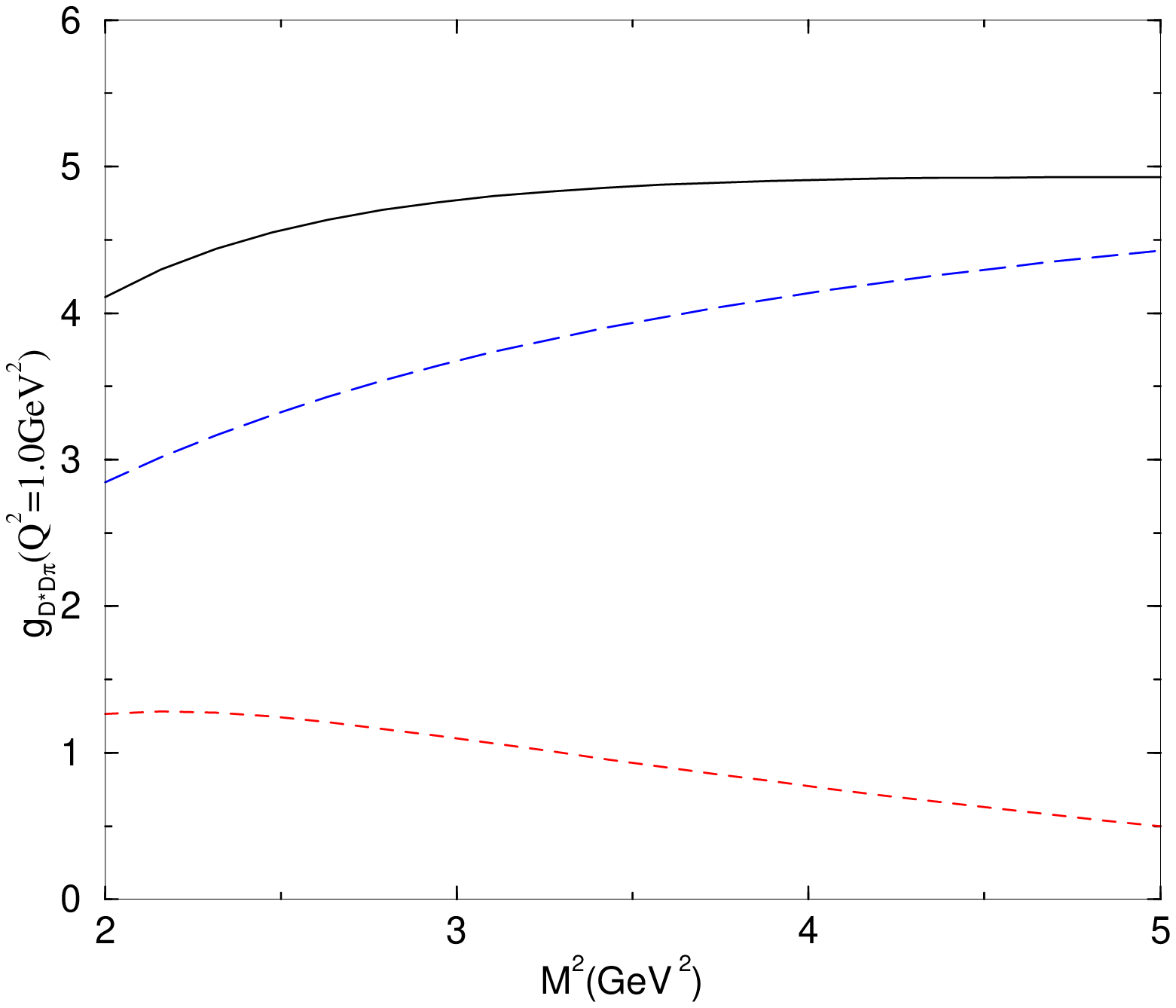}
\caption{$M^2$ dependence of the perturbative (long-dashed line) and 
gluon condensate (dashed line)
contributions to the $D^*D\pi$ form factor at $Q^2=1\,\GeV^2$ (solid line) for
$\Delta_s=\Delta_u=0.5\,\GeV$.}
\end{center}
\end{figure}

\begin{figure} \label{fig2}
\begin{center}
\vskip -1cm
\epsfysize=9.0cm
\epsffile{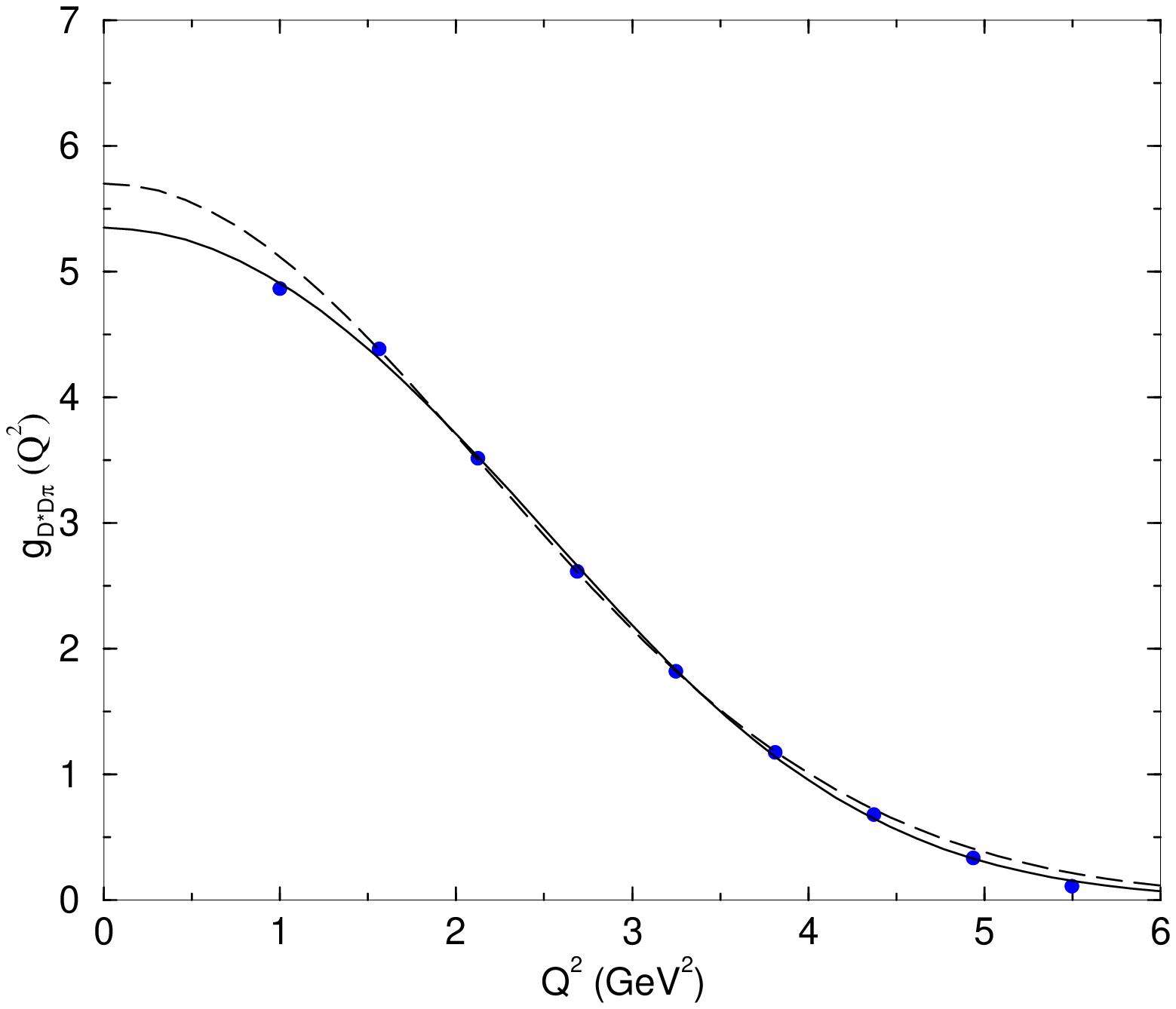}
\caption{Momentum dependence of the $D^*D\pi$ form factor for
$\Delta_s=\Delta_u=0.5\,\GeV$ (dots). The solid and dashed lines give the 
parametrization of the QCDSR results through Eqs.~(\protect\ref{ff})
and (\protect\ref{ga}) respectively.}
\end{center}
\end{figure}

\begin{figure} \label{fig3}
\begin{center}
\epsfysize=9.0cm
\epsffile{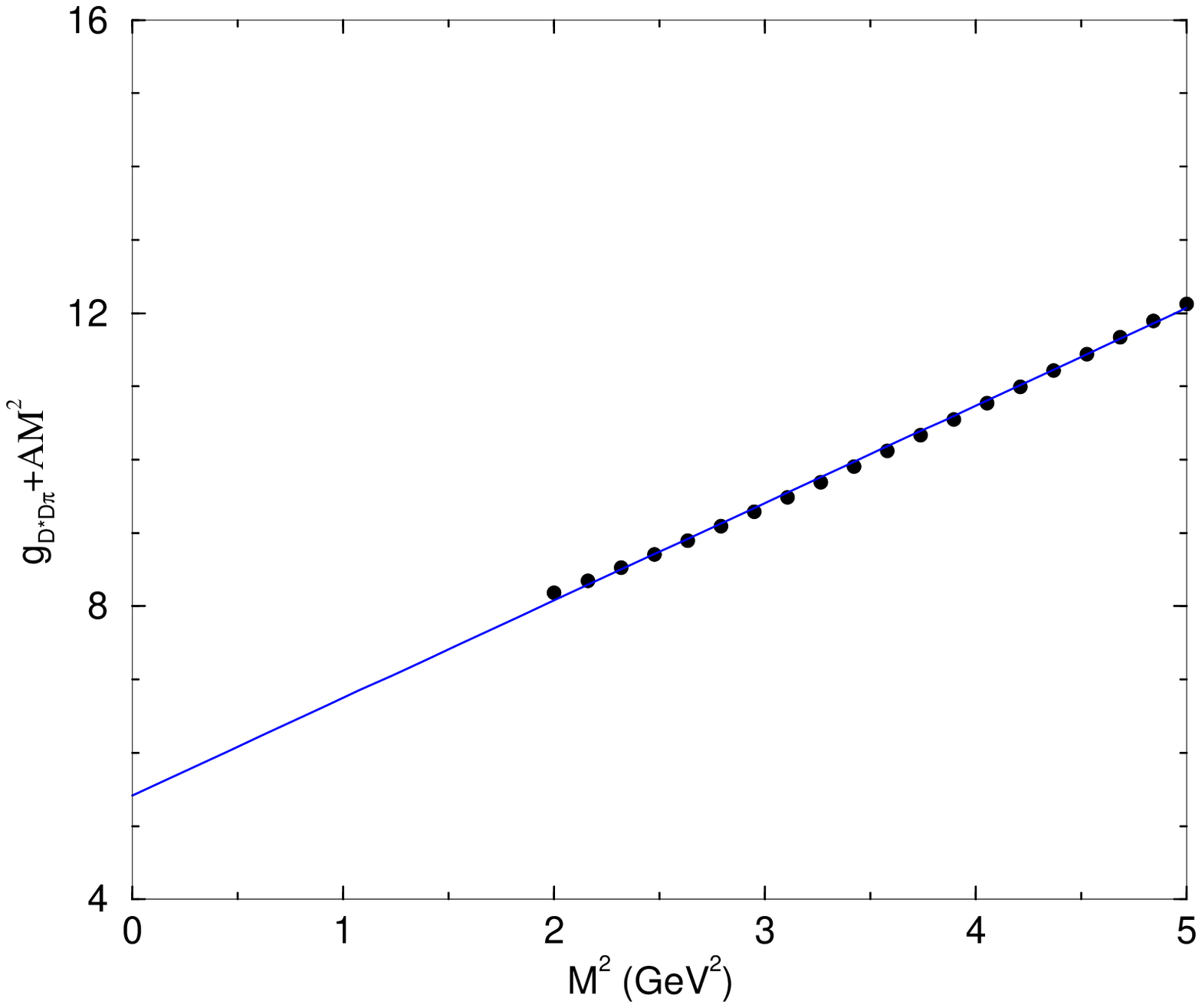}
\caption{$D^*D\pi$ coupling constant as a function of the squared Borel mass
$M^2$ from the QCDSR valid at $Q^2=0$ (dots). The straight line gives
the extrapolation to $M^2=0$.}
\end{center}
\end{figure}

\begin{figure} \label{fig4}
\begin{center}
\vskip -1cm
\epsfysize=9.0cm
\epsffile{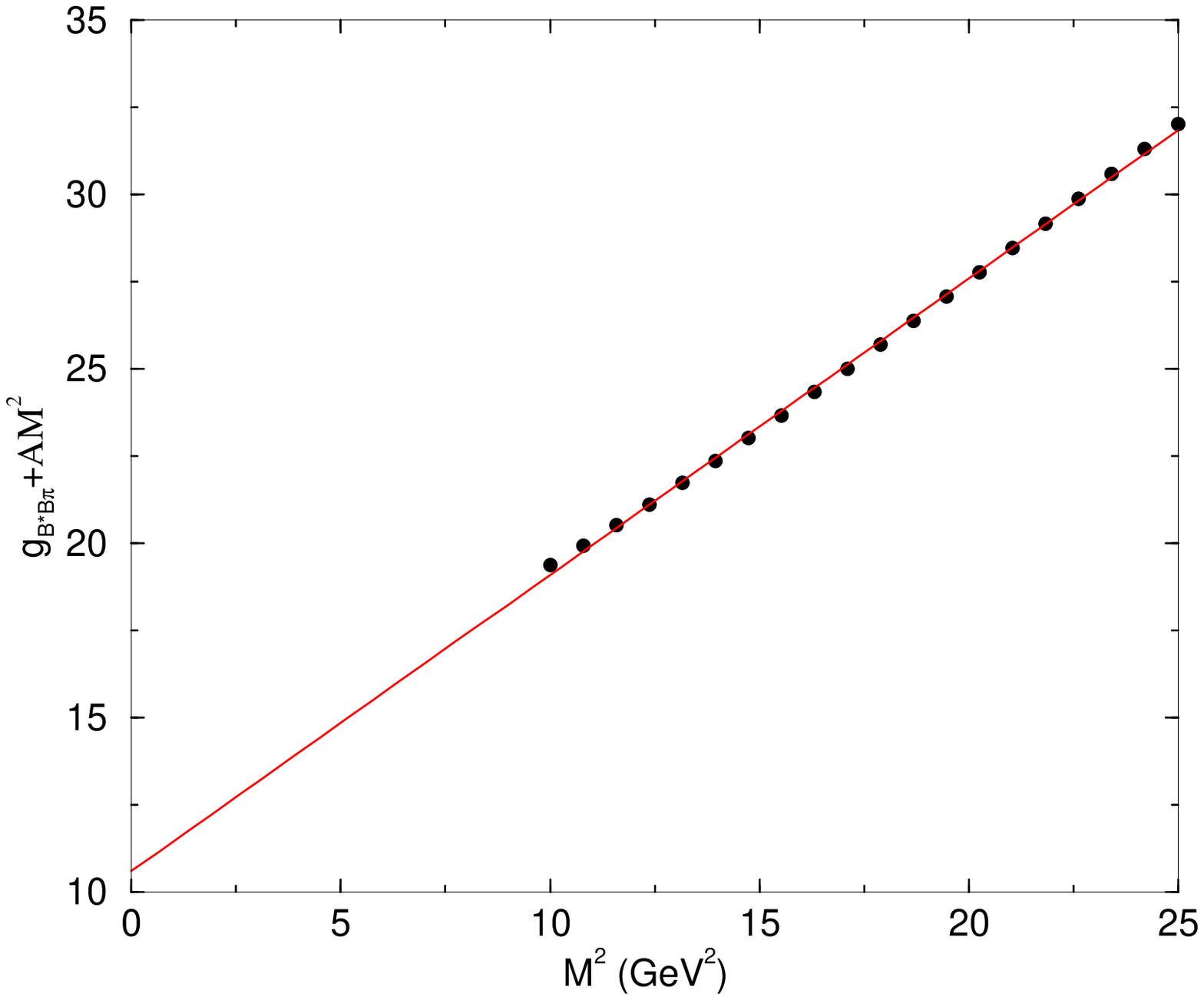}
\caption{$B^*B\pi$ coupling constant as a function of the squared Borel mass
$M^2$ from the QCDSR valid at $Q^2=0$ (dots).The straight line gives
the extrapolation to $M^2=0$.}
\end{center}
\end{figure}

\begin{figure} \label{fig5}
\leavevmode
\begin{center}
\epsfysize=9.0cm
\epsffile{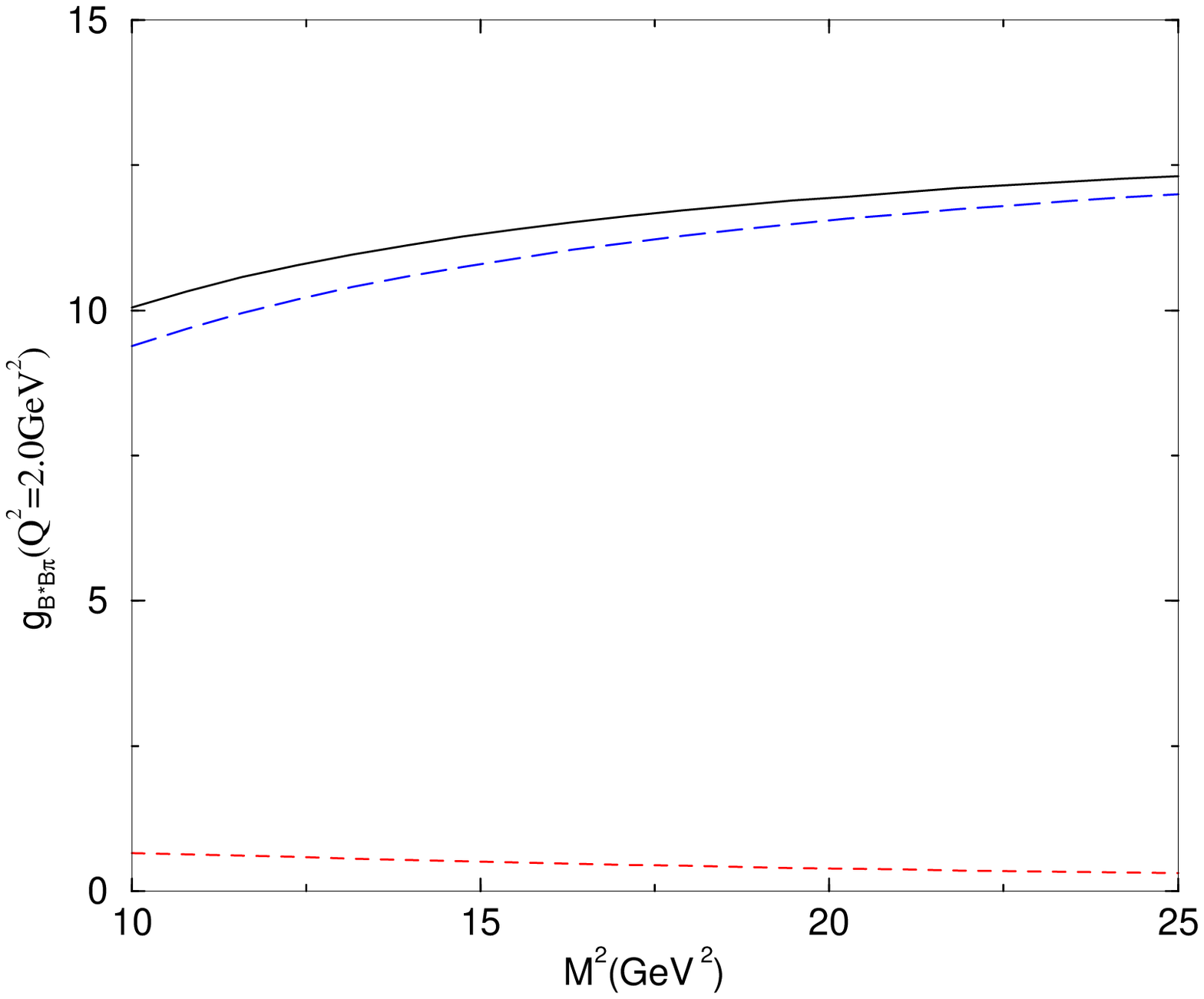}
\caption{$M^2$ dependence of the perturbative (long-dashed line) and 
gluon condensate (dashed line)
contributions to the $B^*B\pi$ form factor at $Q^2=2\,\GeV^2$ (solid line) for
$\Delta_s=\Delta_u=0.5\,\GeV$.}
\end{center}
\end{figure}

\begin{figure} \label{fig6}
\leavevmode
\begin{center}
\vskip -1cm
\epsfysize=9.0cm
\epsffile{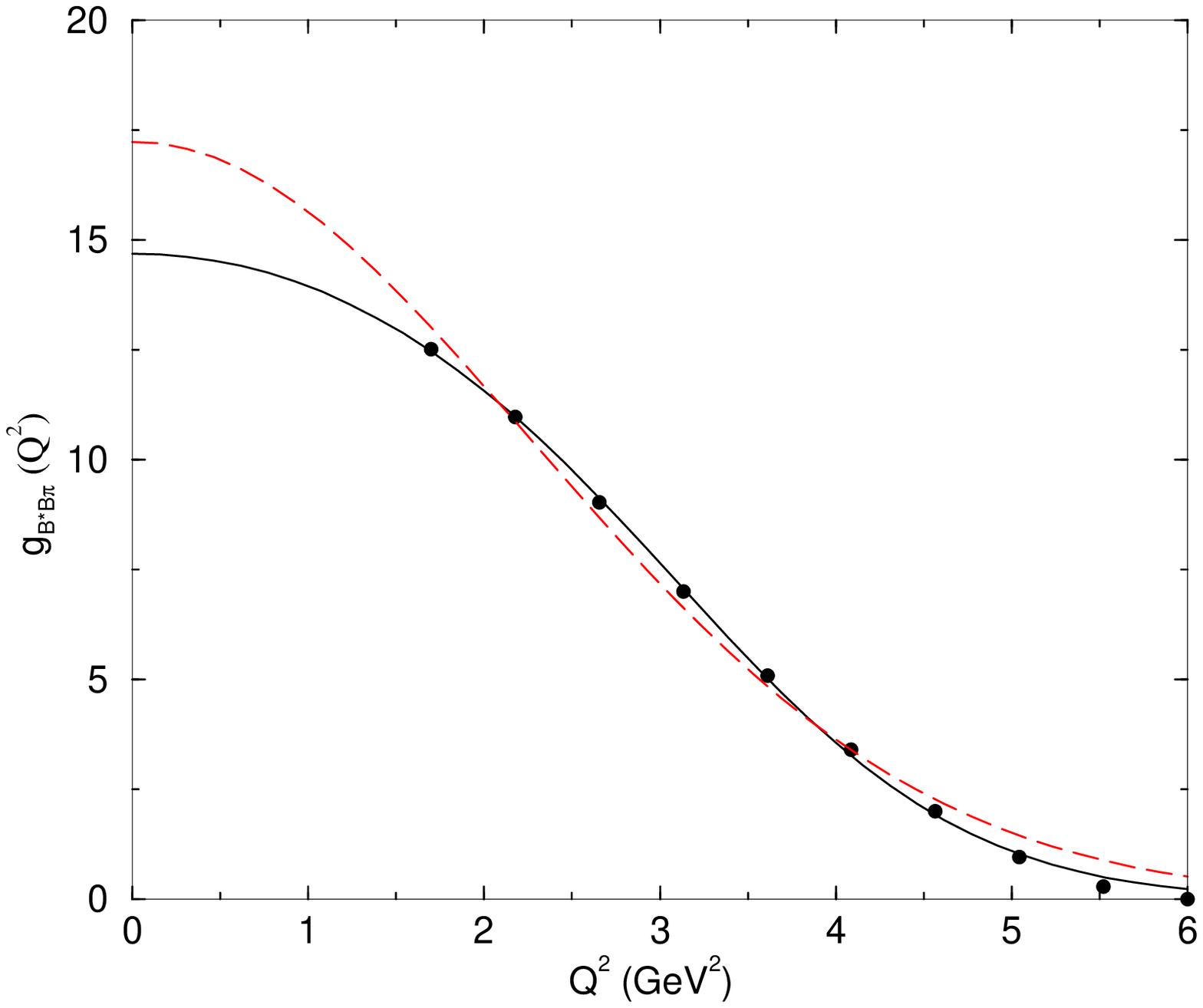}
\caption{Momentum dependence of the $B^*B\pi$ form factor for
$\Delta_s=\Delta_u=0.5\,\GeV$ (dots). The solid and dashed lines give the 
parametrization of the QCDSR results through Eqs.~(\protect\ref{ff})
and (\protect\ref{ga}) respectively.}
\end{center}
\end{figure}

\end{document}